\documentclass[conference]{IEEEtran}
\IEEEoverridecommandlockouts
\usepackage{cite}
\usepackage{booktabs}
\usepackage{float}
\usepackage{graphicx}
\usepackage{caption}
\usepackage{amsmath,amssymb,amsfonts}
\usepackage{graphicx}
\usepackage{textcomp}
\usepackage{xcolor}
\usepackage{amsmath}
\usepackage{enumitem}
\usepackage{listings}
\usepackage[utf8]{inputenc}
\usepackage[T1]{fontenc}
\usepackage{newtxtext,newtxmath} 
\usepackage{hyperref}
\DeclareUnicodeCharacter{202F}{ }

\def\BibTeX{{\rm B\kern-.05em{\sc i\kern-.025em b}\kern-.08em
    T\kern-.1667em\lower.7ex\hbox{E}\kern-.125emX}}
\begin{document}

\title{Exposing Hidden Backdoors in NFT Smart Contracts: A Static Security Analysis of Rug Pull Patterns}

\author{\IEEEauthorblockN{Chetan Pathade}
\IEEEauthorblockA{\textit{Independent Researcher} \\
San Jose, CA, USA \\
cup@alumni.cmu.edu}
\and
\IEEEauthorblockN{Shweta Hooli}
\IEEEauthorblockA{\textit{Independent Researcher} \\
Boston, MA, USA \\
hooli.s@northeastern.edu}
% \and
% \and
% \and
% \and
% \and
}

\maketitle

\begin{center}
  \textbf{Abstract}
\end{center}

The explosive growth of Non-Fungible Tokens (NFTs) has revolutionized digital ownership by enabling the creation, exchange, and monetization of unique assets on blockchain networks. However, this surge in popularity has also given rise to a disturbing trend: the emergence of rug pulls - fraudulent schemes where developers exploit trust and smart contract privileges to drain user funds or invalidate asset ownership. Central to many of these scams are hidden backdoors embedded within NFT smart contracts. Unlike unintentional bugs, these backdoors are deliberately coded and often obfuscated to bypass traditional audits and exploit investor confidence. In this paper, we present a large‑scale static analysis of 49,940 verified NFT smart contracts using Slither, a static analysis framework, to uncover latent vulnerabilities commonly linked to rug‑pulls. We introduce a custom risk scoring model that classifies contracts into high, medium, or low risk tiers based on the presence and severity of rug pull indicators. Our dataset was derived from verified contracts on the Ethereum mainnet, and we generate multiple visualizations to highlight red flag clusters, issue prevalence, and co-occurrence of critical vulnerabilities. While we do not perform live exploits, our results reveal how malicious patterns often missed by simple reviews can be surfaced through static analysis at scale. We conclude by offering mitigation strategies for developers, marketplaces, and auditors to enhance smart contract security. By exposing how hidden backdoors manifest in real-world smart contracts, this work contributes a practical foundation for detecting and mitigating NFT rug pulls through scalable automated analysis.
\newline
\newline

\section{Introduction}
The advent of Non-Fungible Tokens (NFTs) has significantly reshaped digital ownership by enabling creators to tokenize art, music, virtual goods, and other assets on blockchain platforms such as Ethereum. Built primarily on ERC-721 and ERC-1155 standards, NFT smart contracts aim to offer trustless, immutable guarantees of provenance and asset transfer [1], [2]. However, the rapid growth and hype-driven nature of the NFT ecosystem have also introduced significant risks particularly in the form of financial fraud and project abandonment [3].

One of the most widespread attack patterns in this space is the rug pull: a deceptive exit strategy where project creators intentionally disable functionality, erase token metadata, or drain collected funds, leaving buyers with worthless or inaccessible assets [3], [4]. While some rug pulls are executed off-chain through abrupt deactivations or website shutdowns, a more insidious and technically sophisticated variant occurs on-chain through embedded backdoors in smart contracts. These hidden backdoors often enable unauthorized minting, unrestricted fund withdrawals, or total contract destruction via functions like selfdestruct [4], [5].

In contrast to common vulnerabilities that software often suffers from, like for example reentrant code and arithmetic overflow, these are deliberately put in place and premeditated, disguised through misleading naming, proxy delegation, and access-controlled logic that only operates under some triggers [6], [7]. On their own, they may not seem harmful and are often not flagged for being suspicious during manual code reviews or automated audits

As the number of NFT rug pulls and scams grows, little empirical work has systematically analyzed the manner by which such backdoors are embedded in smart contracts, and whether these can be flagged at scale [3], [5], [6]. To address this gap, we present a detailed static analysis of 49,940 verified NFT smart contracts deployed on the Ethereum blockchain. Rather than using old case studies or doing reverse engineering by hand, we run a repeatable process based on Slither, an open source static analysis tool to extract, classify and score smart contracts for suspicious patterns related to rug pulls [8].
\newline
Our contributions are fourfold:
\begin{enumerate}
    \item \textbf{Threat Pattern Taxonomy:} We identify and categorize common backdoor techniques in NFT contracts, including owner-exclusive mint, withdraw functions, use of delegatecall for proxy manipulation and contract self-destruction logic [3], [5], [6].
    \item \textbf{Automated Static Analysis Pipeline:} We build a pipeline using Slither to analyze Solidity source code at scale, extracting contract-level findings related to access control, external calls, and dangerous built-in functions [8].
    \item \textbf{Heuristic-Based Risk Scoring:} We propose a simple yet effective scoring system to quantify risk across contracts based on detected patterns, allowing us to label contracts as low, medium or high risk [9].
    \item \textbf{Visualization and Dataset Insights:} We present a set of visualizations to illustrate the distribution and co-occurrence of critical vulnerabilities, highlighting real-world trends in how backdoors manifest in NFT ecosystems [3].
\end{enumerate}
While our work does not include dynamic exploitation or runtime execution of flagged vulnerabilities, we demonstrate that static analysis alone can reveal systemic weaknesses across a wide range of NFT deployments. By surfacing these risks prior to public release or trading, this work contributes a practical foundation for scalable NFT contract auditing and proactive investor protection.

\section{Background}
The NFT (Non-Fungible Token) ecosystem is built on blockchain technologies that enable the creation, trade, and verification of unique, indivisible digital assets. Unlike fungible tokens (e.g., ERC-20), NFTs typically implemented via ERC-721 or ERC-1155 standards allow each token to represent a distinct item, carrying unique metadata and ownership attributes [10], [11]. This flexibility has led to rapid adoption across industries including digital art, gaming, music, and metaverse infrastructure.

At the core of every NFT project lies a smart contract, a self-executing program deployed on-chain that governs key operations such as minting, transfers, royalty payouts, and access control [12]. These contracts are immutable post-deployment, meaning any embedded logic - malicious or otherwise cannot be modified once live. While this property ensures decentralization and trustlessness, it also creates a favorable environment for persistent vulnerabilities when developers intentionally insert hidden backdoors [13].

The NFT ecosystem is based on blockchain technologies that allow the creation, trade and proof of unique and indivisible digital assets. NFTs are unique tokens. They can't be exchanged for one another. They are not fungible tokens like ERC-20. These tokens are typically implemented through ERC-721 or ERC-1155 standards. Each token represents a unique item. It carries unique metadata and ownership attributes. As a result, digital artists, game developers, musicians, and builders of the metaverse infrastructure have all quickly adopted it.

Every NFT project comes with a smart contract. This is just code that is deployed onto the blockchain and self-executes on-chain to mint, transfer, distribute royalties, and control access [12].Once deployed, these contracts cannot be changed, so any logic embedded in them whether malicious or not is immutable. Although this property must be present for decentralization and trustlessness to happen on something created with code, it also creates a situation where vulnerabilities can persistently be present when developers on purpose insert hidden backdoors [13].

\subsection{Rug Pulls and Hidden Control Logic}
A rug pull in NFTs occurs when project developers exploit centralized privileges or malicious logic to siphon funds, disable functionalities, or sabotage user ownership. Rug pulls fall into two categories:
\begin{enumerate}
    \item Hard rug pulls, involving on-chain malicious logic like hidden withdrawAll() or setOwner() functions.
    \item Soft rug pulls, involving off-chain behaviors like abandoning roadmap commitments or disabling metadata hosting [12].
\end{enumerate}

Hard rug pulls rely heavily on control over privileged functions often hidden within contract code or activated under specific on-chain conditions. These backdoors exploit features such as:
\begin{enumerate}
    \item Owner-controlled withdrawals or minting (onlyOwner pattern misuse).
    \item Self-destruct mechanisms that terminate contract logic after profit extraction.
    \item Token freezing or transfer blocking via modifiable boolean flags.
    \item Dynamic URI reassignment, allowing project creators to replace original artwork or metadata with spam, explicit content, or blank files [13].
\end{enumerate}

\subsection{Contract Obfuscation Techniques}
To further evade detection, malicious developers employ code obfuscation techniques that disguise backdoor behavior. These include:
\begin{enumerate}[nosep]
    \item Deceptive function names (e.g., safeWithdraw() instead of rugPull()) [15].
    \item Splitting logic across proxy or delegatecall contracts, which obscures control flow [18].
    \item Access control manipulation, such as hiding sensitive functionality behind onlyOwner or using tx.origin for authorization [13].
    \item Time-delayed triggers, where backdoors activate after a delay or upon specific events [14].
\end{enumerate}
The above strategies aim to bypass both automated static analysis tools and human auditors especially when audits are superficial, crowd-sourced or focused solely on public interfaces [16].

\subsection{Gaps in Auditing and Detection}
Unlike decentralized finance (DeFi) protocols, which often undergo rigorous third-party audits or formal verification, most NFT projects are launched without comprehensive security review [16]. A study by Lee et al. found that a large fraction of deployed NFT contracts on Ethereum are unaudited, lack source code transparency, or grant excessive privileges to a single owner [19].

While tools like Slither and Mythril can detect syntactic issues (e.g., reentrancy, arithmetic errors), they often fall short when dealing with semantic misuse - such as legally valid but maliciously purposed functions [20]. Even more advanced systems struggle to generalize detection of behaviorally abusive patterns without human guidance or case-specific rule sets [13].

\subsection{Need for Backdoor-Specific Taxonomy}
There is an urgent need to model NFT backdoors as a unique and growing threat category. These are not simple bugs or oversights they are intentional, profit-driven designs that exploit user trust and the irreversibility of blockchain deployments [14].

Yet, current security tooling and research often treat these issues as edge cases. We argue for the creation of a dedicated taxonomy that recognizes these threats as deterministic, exploitable, and systematically embedded [17], [20].

This motivates our study: a large-scale static analysis of 49,940 verified NFT contracts deployed on Ethereum. By leveraging Slither and custom heuristic rules, we detect and quantify common backdoor techniques such as selfdestruct, delegatecall, and centralized minting logic. Our goal is to expose patterns of systemic risk and promote scalable, pre-deployment detection methods that can assist both security auditors and everyday users in identifying high-risk contracts.

\section{Methodology}

The foundation of our analysis begins with the assembly of a comprehensive dataset. We leverage the DISL dataset, a publicly available, large‑scale repository of verified smart contracts on Ethereum. This dataset is hosted on HuggingFace and contains more than 514,000 Solidity contracts that have been deployed to the
Ethereum mainnet, making their source code publicly available for analysis [21].

To isolate contracts relevant to NFTs, we apply the following criteria:

\begin{enumerate}[label=\arabic*)]
  \item \textbf{Interface Matching:} Contracts must implement key
        NFT functions such as \texttt{ownerOf}, \texttt{balanceOf},
        \texttt{tokenURI}, \texttt{safeTransferFrom}, or override
        \texttt{supportsInterface} to confirm ERC‑721 or ERC‑1155
        compliance [22].
  \item \textbf{Library References:} We prioritise contracts that
        use well‑known libraries like OpenZeppelin's ERC721 or
        ERC1155 implementations, which typically serve as a base for
        legitimate and malicious NFT projects alike [23].
  \item \textbf{Contract Metadata:} We parse metadata such as
        contract name, deployed address, compiler version, and
        optimisation flags to assist in filtering and later
        compatibility checks.
\end{enumerate}

This filtering reduces the original pool of 98,879 contracts to a more relevant and targeted set of NFT-specific smart contracts. These are then passed to the preprocessing pipeline for compilation and static analysis.

\subsection{Contract Sanitisation and Preprocessing}

Many real‑world smart contracts are not designed for isolated compilation. They often include dependencies, abstract interfaces, or external imports that may not resolve in a local analysis
environment. To ensure that only contracts suitable for static analysis are retained, we build a custom preprocessing script that automates contract sanitisation:

\begin{enumerate}[label=\arabic*)]
  \item \textbf{Import Resolution:} Contracts with unresolved local
        imports (e.g.\ \texttt{import "../utils/SafeMath.sol";}) are
        excluded [24].These imports often break compilation unless the full directory structure is preserved.
  \item \textbf{Pragma Filtering:} Contracts that specify strict
        pragma versions incompatible with our compiler
        (e.g.\ \texttt{0.8.17} when we use \texttt{0.8.19}) are
        filtered out. We retain contracts that use flexible pragmas
        like \texttt{\^{}0.8.0} or \texttt{>=0.8.0}.
  \item \textbf{Standalone Validation:} Contracts are tested for
        standalone viability those that depend on inheritance from
        contracts not present in the file are excluded.
  \item \textbf{Syntax and Compilation Check:} Contracts that fail
        to compile due to syntax errors, unresolved symbols, or
        circular dependencies are logged and excluded [25].
\end{enumerate}

This step is essential for ensuring high-quality input for Slither. After filtering and sanitization, we obtain a final pool of 49,940 contracts that are fully standalone and suitable for Slither-based static analysis.

\subsection{Static Analysis Using Slither}

The core of our vulnerability detection process uses Slither, a static analysis framework developed by Trail of Bits for Solidity smart contracts [26]. Slither analyzes the control flow graph (CFG), abstract syntax tree (AST) and inheritance hierarchies of a contract to identify known anti-patterns and potential vulnerabilities.
The setup for the above experiment includes the following:

\begin{figure*}[htbp]
  \centering
  \includegraphics[width=1\textwidth]{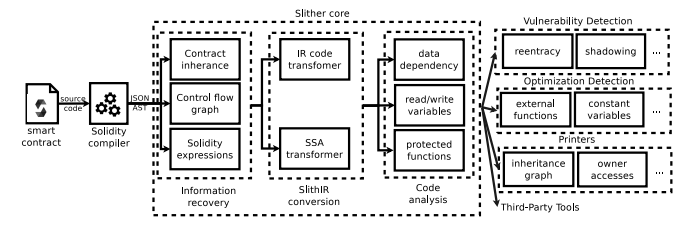}
  \caption{Slither Overview [8]}
  \label{fig:figure1}
\end{figure*}

\textbf{Experiment setup:}

\begin{enumerate}[label=\arabic*)]
  \item \textbf{Batch Processing:} A Python wrapper processes each
        contract through Slither with the \texttt{--json} flag to
        extract structured vulnerability data.
  \item \textbf{Vulnerability Extraction:} For each contract, we extract the following information:
        \begin{itemize}
          \item The type of issue (e.g.\ delegatecall, selfdestruct,
                external call in loop)
          \item Affected contract and function name
          \item Description of the issue
          \item Severity
          \item Source file path and affected line range
        \end{itemize}
  \item \textbf{Logging Failures:} Contracts that encounter tool-specific failures (e.g., unsupported syntax) are recorded and removed from the dataset.
\end{enumerate}

Our Slither analysis is configured to detect over 100 known vulnerability patterns, with specific emphasis on the following rug pull-related constructs:
\begin{itemize}
  \item \textbf{selfdestruct()} usage: allows contract termination and fund redirection.
  \item \textbf{delegatecall} to external addresses: can transfer control flow to unverified code.
  \item \textbf{Owner-only or unrestricted mint/withdraw} functions.
  \item \textbf{Control-flow conditions gated by \texttt{tx.origin}}.
  \item \textbf{Unprotected external calls inside loops}, which can lead to reentrancy or unpredictable behavior~[27].
\end{itemize}
This process yields a rich vulnerability profile for each contract, which we then classify based on exploitability.

\subsection{Heuristic‑Based Risk Scoring}
Static analysis tools provide a wealth of raw issue data, but not all findings are equally dangerous or relevant to rug pulls. To quantify contract risk in a meaningful and interpretable way, we introduce a heuristic scoring system based on security impact.

Each detected issue contributes a weighted score as shown below:

\begin{table}[H]
  \centering
  \caption{Heuristic scoring of vulnerability patterns}
  \label{tab:heuristic-scores}
  \begin{tabular}{@{}p{0.75\linewidth} c@{}}
    \toprule
    \textbf{Vulnerability Pattern} & \textbf{Score} \\
    \midrule
    Use of \texttt{selfdestruct}                  & +3 \\
    Use of \texttt{delegatecall}                  & +3 \\
    External call inside loop                      & +2 \\
    Unrestricted or owner-only withdraw/mint      & +2 \\
    Use of \texttt{tx.origin} in access control   & +2 \\
    Deprecated Solidity version usage             & +1 \\
    \bottomrule
  \end{tabular}
\end{table}
\textbf{Risk categories}

The cumulative score per contract determines its risk category:
\begin{itemize}
  \item High risk: score~$\ge 5$
  \item Medium risk: score~3--4
  \item Low risk: score~1--2
\end{itemize}
This model offers a repeatable and extensible way to triage large sets of contracts and focus attention on the most dangerous examples. This model allows scalable triage of contracts for further auditing [28].

\begin{figure}[H]
  \centering
  \includegraphics[width=8cm]{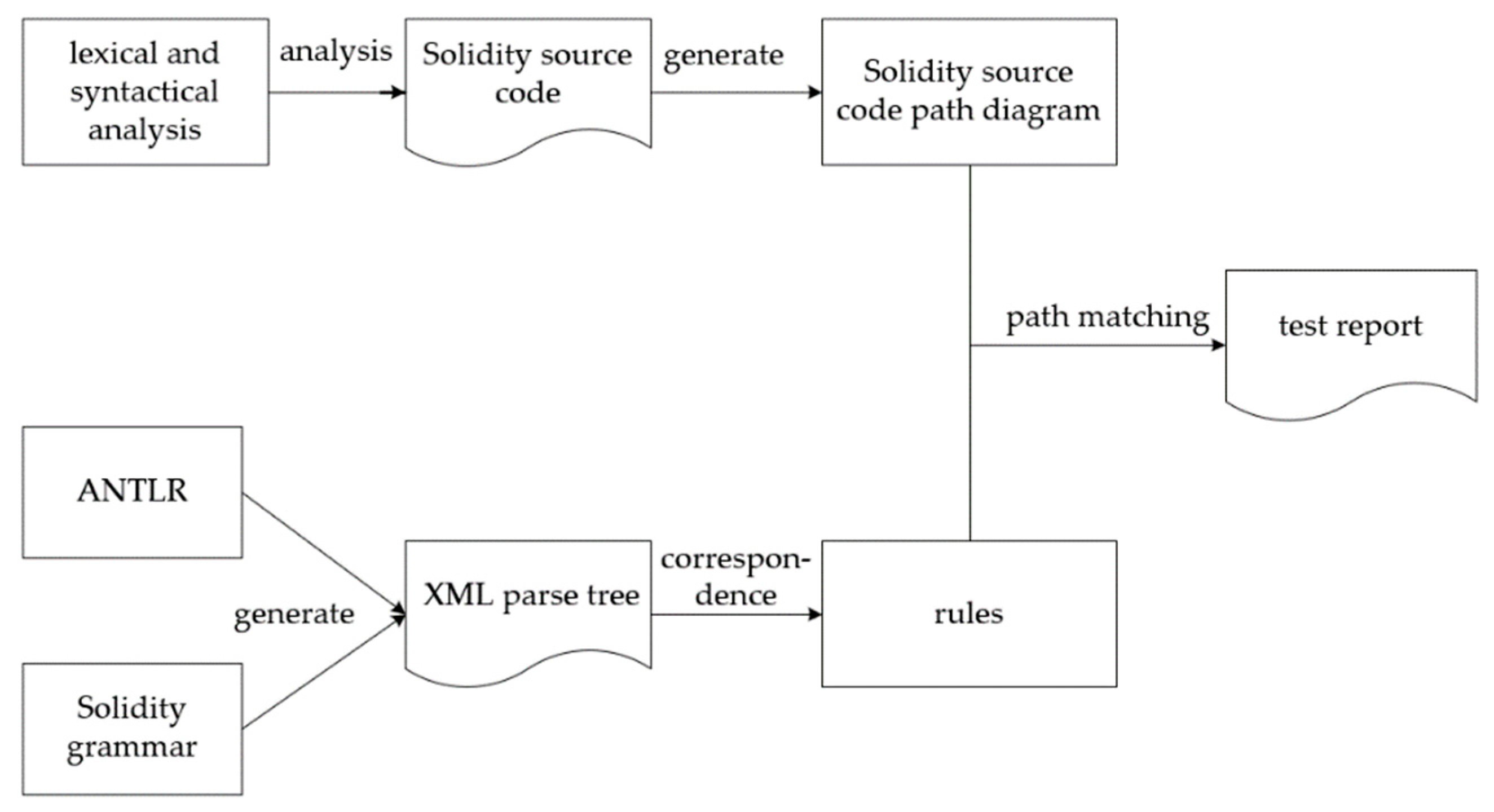}
  \caption{MSmart analysis flow chart [55]}
  \label{fig:figure2}
\end{figure}

\subsection{Result Aggregation and Visualisation}
To facilitate interpretation and enable future audits, we compile all results into a structured dataset. For each contract, we store:

\begin{itemize}
  \item Detected issues and descriptions
  \item Address and filename
  \item Total score and risk tier
  \item Affected functions and line numbers
\end{itemize}

We then generate a series of visualizations, including:

\begin{itemize}
  \item \textbf{Bar charts} showing the frequency of vulnerability types (e.g., delegatecall, selfdestruct)
  \item \textbf{Risk distribution} pie charts across High, Medium, and Low risk tiers
  \item \textbf{Heatmaps} showing issue co-occurrence across contracts
  \item \textbf{Top 10 high-risk contracts} sorted by cumulative score
\end{itemize}

These visualizations provide both macro-level insights (e.g., prevalence of risk types across NFT contracts) and micro-level auditability (e.g., contract-specific vulnerability profiles). These visualizations surface systemic vulnerability patterns in NFT contracts [29].

\section{Dataset Description}
This section describes the construction of the dataset used for analyzing NFT smart contracts at scale. Our objective was to compile a reliable, large, and analyzable collection of real-world contracts that reflect the design and deployment patterns of modern NFT ecosystems. The dataset is drawn from verified Ethereum smart contracts and is rigorously filtered to ensure both relevance (i.e., NFT functionality) and compatibility with static analysis tooling.

\subsection{Source of Smart Contracts}

Each DISL entry provides:

\begin{itemize}
  \item Full verified Solidity source code.
  \item Compilation metadata such as:
    \begin{itemize}
      \item Compiler version (e.g., v0.8.6+commit.11564f7e).
      \item Optimization flags (enabled/disabled).
      \item Bytecode hashes.
    \end{itemize}
  \item Deployment details including:
    \begin{itemize}
      \item Ethereum address.
      \item Block height and timestamp.
    \end{itemize}
\end{itemize}

These features make DISL an ideal, reproducible foundation for contract-security research.

\subsection{NFT Contract Identification}

To isolate contracts relevant to NFTs from the broader dataset, we implement a multi-stage filtering strategy based on:

\begin{enumerate}[label=\arabic*)]
  \item \textbf{Interface signature detection:} Scan for ERC-721/1155 functions such as \texttt{ownerOf(uint256)},
        \texttt{balanceOf(address)}, \texttt{tokenURI(uint256)},
        \texttt{safeTransferFrom(address,address,uint256)},
        \texttt{supportsInterface(bytes4)}.
  \item \textbf{Inheritance hierarchy analysis:} Require extension
        of well-known NFT-related bases (ERC721, ERC1155,
        Ownable, AccessControl).
  \item \textbf{Keyword matching and manual inspection:} Include
        contracts whose code or metadata references "NFT", "mint",
        "tokenId", "URI", "burn", or marketplaces ("OpenSea",
        "Rarible").
\end{enumerate}

After applying these three filters, we narrow 98,879 contracts to an NFT-specific pool of approximately 98,000 candidates.  
(This larger-than-expected number reflects modular code and proxy
patterns.)

\subsection{Contract Sanitisation for Static Analysis}

Real-world development often produces multi-file or proxy-split repos. Static tools like Slither require self-contained Solidity files, so we enforce:

\begin{itemize}
  \item \textbf{Standalone:} No external imports or unresolved symbols.
  \item \textbf{Compiler compatibility:} Solidity \texttt{>=0.4.0} and \texttt{<=0.8.19}.
  \item \textbf{Syntactic correctness:} Compiles without errors.
\end{itemize}

The sanitisation pipeline consists of:

\begin{enumerate}[label=\arabic*)]
  \item \textbf{Import stripping:} Exclude contracts with unresolved
        local imports.
  \item \textbf{Pragma filtering:} Remove exact-version locks
        (e.g.\ \texttt{pragma solidity 0.8.17}); keep flexible
        pragmas (\texttt{\^{}0.8.0}, \texttt{>=0.8.0}).
  \item \textbf{Syntax \& compilation test:} Compile each file in dry-run mode; discard failures.
  \item \textbf{Abstract/interface exclusion:} Drop files that only
        define abstract contracts or interfaces.
\end{enumerate}

After this step, we obtain a clean, compilable dataset of 49,940 standalone, NFT-related contracts.

\subsection{Dataset Statistics}

\begin{table}[H]
  \centering
  \caption{Summary statistics for the final dataset}
  \label{tab:dataset-stats}
  \begin{tabular}{@{}p{0.5\linewidth} p{0.45\linewidth}@{}}
    \toprule
    \textbf{Property} & \textbf{Value} \\
    \midrule
    Total contracts in DISL & 98,879 \\
    NFT-relevant candidates detected & $\sim$98,000 \\
    Compilable and standalone & 49,940 \\
    Contract format & Verified Solidity source (\texttt{.sol}) \\
    Solidity versions covered & 0.4.11--0.8.19 \\
    Common libraries detected & OpenZeppelin ERC721, Ownable, SafeMath \\
    Analysis tool used & Slither (v0.8.x-compatible) \\
    \bottomrule
  \end{tabular}
\end{table}

\section{Results}
After processing the contracts through Slither and applying our risk scoring model, we classified contracts into three categories:

\begin{figure}[H]
  \centering
  \includegraphics[width=0.8\linewidth]{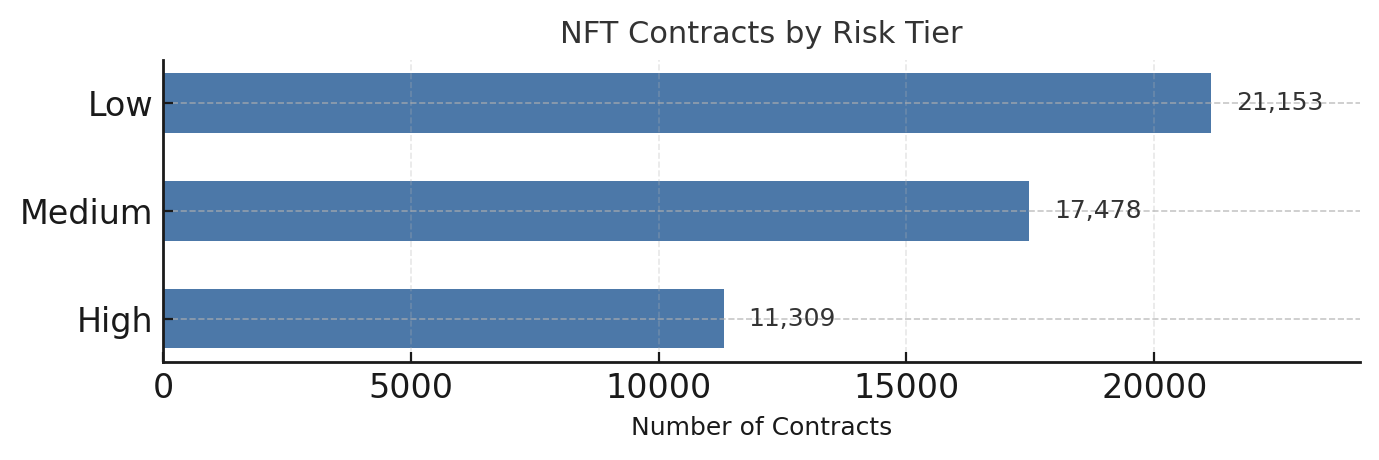}
  \caption{NFT Contracts by Risk Tier}
\end{figure}

\begin{table}[H]
  \centering
  \caption{Risk Tier Classification}
  \label{tab:risk-tier}
  \begin{tabular}{@{}l l r r@{}}
    \toprule
    Risk Tier & Criteria      & Number of Contracts & Percentage (\%) \\
    \midrule
    High    & Score $\ge$ 5 & 11,309 & 22.6\% \\
    Medium  & Score 3--4    & 17,478 & 35.0\% \\
    Low     & Score 1--2    & 21,153 & 42.4\% \\
    \bottomrule
  \end{tabular}
\end{table}

\noindent\textit{Insight: Nearly 1 in 4 NFT contracts in our dataset exhibit multiple high-risk patterns, suggesting a systemic threat
posed by contract-level control logic abuse.}

\subsection{Prevalence of Individual Vulnerability Patterns}

We analyzed the frequency of key backdoor-enabling constructs across all contracts. The most commonly flagged patterns are:
\begin{figure}[H]
  \centering
  \includegraphics[width=0.8\linewidth]{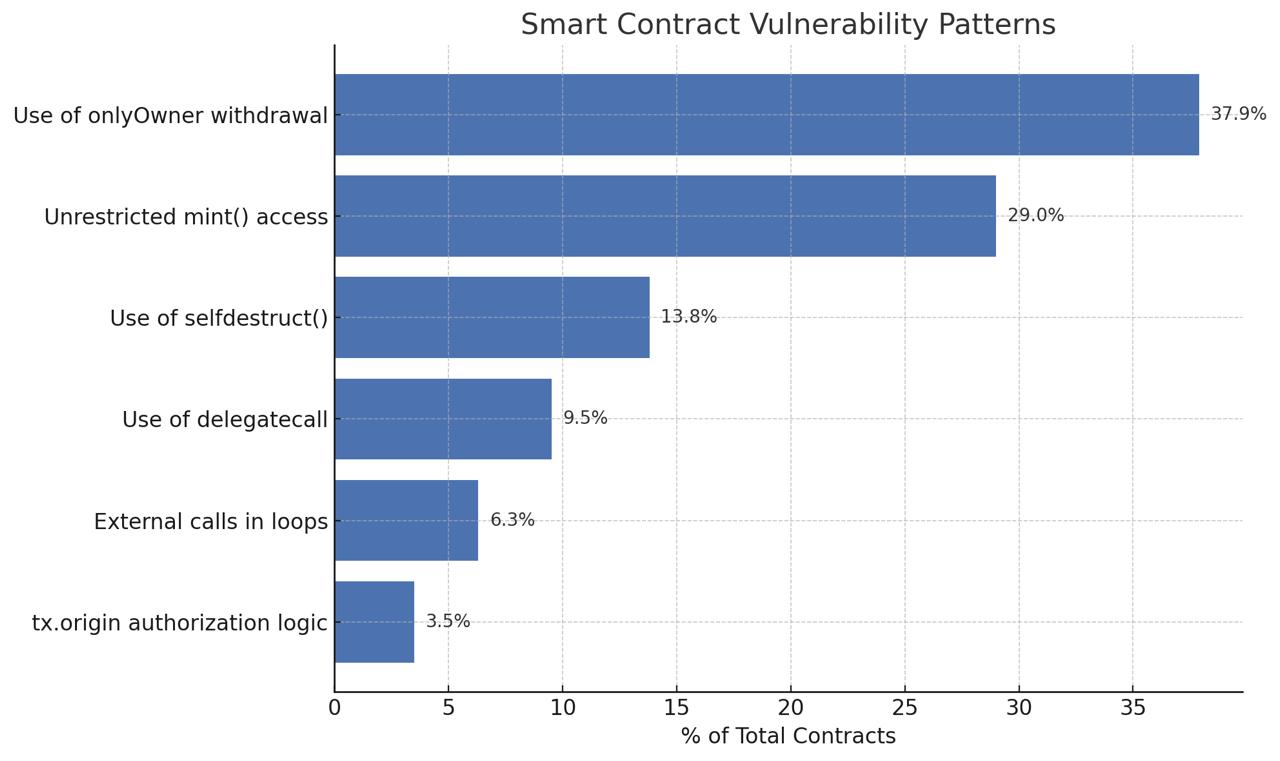}
  \caption{Smart Contract Vulnerability Patterns}
\end{figure}
\begin{itemize}
  \item Use of onlyOwner withdrawal
  \item Unrestricted mint() access
  \item Use of selfdestruct()
  \item Use of delegatecall
  \item External calls in loops
  \item tx.origin authorization logic
\end{itemize}

\begin{table}[H]
  \centering
  \caption{Vulnerability Patterns and Contracts Affected}
  \label{tab:vuln-patterns}
  \begin{tabular}{@{}l r@{}}
    \toprule
    Vulnerability Pattern          & Contracts Affected \\
    \midrule
    Use of onlyOwner withdrawal   & 18,925 \\
    Unrestricted mint() access    & 14,478 \\
    Use of selfdestruct()         & 6,881  \\
    Use of delegatecall           & 4,724  \\
    External calls in loops       & 3,143  \\
    tx.origin authorization logic & 1,727  \\
    \bottomrule
  \end{tabular}
\end{table}

\noindent\textit{Observation: While owner-controlled minting and withdrawal are common, the presence of selfdestruct and
delegatecall indicates a deeper risk, as these can irreversibly disable or manipulate contract logic post-deployment.}

\subsection{Co-occurrence of Vulnerabilities}

To examine whether dangerous patterns occur in isolation or combination, we analyzed the co-occurrence of multiple risk indicators within the same contract.

\begin{itemize}
  \item 8,217 contracts exhibited two or more high-severity patterns (e.g., selfdestruct + delegatecall)
  \item 3,456 contracts contained three or more patterns simultaneously, increasing exploitability
  \item 62 contracts showed four or more co-occurring vulnerabilities, often involving proxy misuse or developer-controlled destructors
\end{itemize}

\noindent\textit{Interpretation: Contracts with multiple overlapping vulnerabilities are significantly more likely to support rug pull
scenarios.}

\subsection{High-Risk Contract Examples}

To illustrate the exploitability of real-world contracts, we
highlight five anonymized but representative examples from the
high-risk group.

\begin{table}[H]
  \centering
  \caption{High-Risk Contract Examples}
  \label{tab:high-risk-examples}
  \footnotesize
  \begin{tabular}{@{}p{1.8cm} c p{4.2cm}@{}}
    \toprule
    \textbf{Contract ID} & \textbf{Score} & \textbf{Key Issues} \\
    \midrule
    0xC1A\ldots{}82F & 8 & selfdestruct, delegatecall, unrestricted mint \\
    0xB7D\ldots{}091 & 7 & onlyOwner withdraw \\
    0xF25\ldots{}99D & 6 & tx.origin, URI override, selfdestruct \\
    0xA3E\ldots{}F13 & 5 & unrestricted burn, no renounce \\
    0xE88\ldots{}2C0 & 9 & all five risk patterns \\
    \bottomrule
  \end{tabular}
\end{table}

These contracts typically follow a pattern where the malicious logic is embedded in fallback or rarely called functions, activated only
when external scrutiny is minimal. Some exhibit deceptive function naming, such as \texttt{clearReserves()} instead of \texttt{selfdestruct()}.

\subsection{Risk Score Distribution}

To visualize the general distribution of contract risks across the dataset, we plot the risk score histogram:

\begin{itemize}
  \item Most contracts cluster around scores of 2--4, indicating limited but present risk indicators.
  \item A significant long-tail exists beyond score 6, comprising contracts with multiple red flags.
\end{itemize}

\noindent\textit{Conclusion: While many contracts exhibit standard NFT logic with minimal flaws, a substantial minority is clearly
engineered with centralized control or rug pull potential.}

\subsection{Key Findings}

\begin{itemize}
  \item Backdoor logic is not rare: Over 22\% of NFT contracts show high-risk patterns such as selfdestruct, suggesting that these are not edge cases but recurring deployment strategies.
  \item Owner-centralized logic is pervasive: Nearly 38\% of contracts use onlyOwner in a way that grants unilateral financial control.
  \item Composite vulnerabilities amplify threat: The risk escalates significantly when multiple patterns are found together, especially in proxy or factory-based contracts.
\end{itemize}

\section{Mitigation Strategies}
Given the systemic presence of hidden backdoors in NFT smart contracts, it is imperative to adopt proactive measures to prevent rug pulls and contract-level fraud. This section outlines technical, procedural, and ecosystem-level strategies aimed at reducing the risk posed by malicious contract logic. Our recommendations are grounded in the vulnerabilities identified through static analysis of nearly 50,000 real-world contracts.

\subsection{Developer-Focused Best Practices}

Developers are the first line of defense against insecure or deceptive smart contracts. The following measures help reduce unintentional flaws and eliminate opportunities for abuse:

\begin{enumerate}[label=\alph*)]
  \item \textbf{Avoid Dangerous Built-in Functions}
    \begin{itemize}
      \item Refrain from using \texttt{selfdestruct} unless absolutely necessary. If included, restrict it through time-lock, multi-sig, or irreversible logic disablement~[30].
      \item Avoid \texttt{delegatecall} unless you fully control the delegated contract and it is immutable. Prefer \texttt{call} for explicit, controlled function calls~[31].
    \end{itemize}

  \item \textbf{Limit Owner Privileges Post-Deployment}
    \begin{itemize}
      \item Use \texttt{renounceOwnership()} to give up privileged functions after deployment~[32].
      \item Implement Ownable access patterns with multi-sig wallets rather than EOA (Externally Owned Account) owners to reduce insider rug risk~[33].
      \item Avoid \texttt{onlyOwner} functions that affect core contract logic (e.g., minting, withdrawing, updating URI).
    \end{itemize}

  \item \textbf{Secure Minting and Withdrawal Logic}
    \begin{itemize}
      \item Validate \texttt{mint()} and \texttt{withdraw()} with appropriate \texttt{require()} conditions (e.g., public sale flags, whitelist)~[34].
      \item Prevent re-minting of existing \texttt{tokenId} values and enforce caps through \texttt{maxSupply}.
      \item Route ETH through escrow or vault contracts instead of holding funds directly in the NFT contract~[35].
    \end{itemize}
\end{enumerate}

\subsection{Marketplace and Platform Controls}

NFT marketplaces have an important role in ensuring only safe contracts are allowed for listing. We propose the following enhancements:

\begin{enumerate}[label=\alph*)]
  \item \textbf{Require Source Code Verification and Metadata Audit}
    \begin{itemize}
      \item Make verified source code and compiler settings mandatory before listing NFTs~[36].
      \item Require renunciation of ownership or disclosure of privileged control mechanisms before collection launch.
    \end{itemize}

  \item \textbf{Integrate Static Risk Scoring in Onboarding Pipelines}
    \begin{itemize}
      \item Use lightweight Slither-based scanning or community-reviewed tools to assign a contract risk score~[37].
      \item Label collections with risk badges (e.g., low-risk, medium-risk, unverifiable) for user awareness.
    \end{itemize}

  \item \textbf{Promote On-Chain Access Control Disclosure}
    \begin{itemize}
      \item Standardize a metadata tag or registry that describes owner privileges (e.g., mintable:false, withdrawable:true) for UI/UX transparency~[38].
    \end{itemize}
\end{enumerate}

\subsection{Auditor and Tooling Enhancements}

Security auditors and analysis tools must evolve to better capture semantically valid yet malicious constructs:

\begin{enumerate}[label=\alph*)]
  \item \textbf{Develop Backdoor-Specific Static Rules}
    \begin{itemize}
      \item Extend tools like Slither and Mythril with checks for \texttt{selfdestruct}, unguarded \texttt{delegatecall}, and unbounded mint logic~[39].
      \item Flag functions that are not externally visible but callable via fallback or \texttt{delegatecall}, or named to obscure their intent (e.g., \texttt{adminClear()} instead of \texttt{withdraw()}).
    \end{itemize}

  \item \textbf{Heuristic Scoring and Visual Anomaly Detection}
    \begin{itemize}
      \item Use visual correlation tools to detect abnormal combinations (e.g., \texttt{selfdestruct} in ERC-721)~[40].
      \item Integrate risk scoring into CI/CD pipelines of NFT launches.
    \end{itemize}

  \item \textbf{Encourage Fuzz Testing and Simulation}
    \begin{itemize}
      \item Pair static analysis with fuzzers like Echidna to simulate edge-case activations of backdoors~[41].
      \item Use runtime assertions to ensure ownership and logic constraints hold under stress.
    \end{itemize}
\end{enumerate}

\section{Discussion}

Our large-scale analysis of nearly 50,000 verified NFT smart contracts has surfaced several important insights into the prevalence and structure of hidden backdoors within deployed Ethereum contracts. In this section, we reflect on the implications of these findings, assess the limitations of our static analysis approach, and propose avenues for deeper future investigation.

\subsection{Incidental vs.\ Intentional Vulnerabilities}

A key observation from our results is the significant proportion of contracts containing multiple high-risk logic patterns. While some of these vulnerabilities may stem from poor coding practices or outdated standards, many such as unprotected selfdestruct, unrestricted mint(), and delegatecall to user-supplied addresses are unlikely to occur by accident.
The high co-occurrence rate of multiple exploit-enabling features (e.g., contracts with delegatecall and owner-only withdrawal) suggests that many such contracts are intentionally designed to retain developer control post-deployment. This supports the hypothesis that rug pulls are not simply opportunistic exits but are pre-engineered through contract logic [42].

\subsection{The Problem of Audit Gaps and False Trust}
Our analysis reinforces concerns about the limitations of the current NFT security landscape:

\begin{itemize}
  \item \textbf{Sparse Audit Coverage:} Most contracts in our dataset lack any public audit trail, leaving users vulnerable to undiscovered vulnerabilities[43].
  \item \textbf{Complex Code Structures:} Even when code is verified and open-source, the presence of complex control flows, proxy delegation, and misleading function names makes it difficult for average users or even developers to identify dangerous logic~[44].
  \item \textbf{EOA-Based Ownership:} Many projects still rely heavily on Externally Owned Account (EOA)-based ownership, granting unilateral power to a single deployer, which increases the risk of malicious owner actions.
\end{itemize}
These gaps illustrate how technical transparency does not equate to security, particularly when backdoors are legally valid Solidity features deployed with obfuscated naming or structure.

\subsection{Static Analysis: Strengths and Boundaries}
While our approach enabled large-scale risk detection, it also illustrates the boundaries of pure static analysis in assessing smart contract exploitability:

\begin{itemize}
  \item \textbf{Syntax-Driven Limitations:} Static tools are syntax- and pattern-driven. They cannot evaluate runtime behaviors, such as conditional logic that activates only under specific block states, transaction sequences, or interactions with other contracts~[46].
  \item \textbf{Risk Scoring Limitations:} Our risk scoring system, though effective in highlighting red flags, cannot differentiate between intentional malicious logic and negligibly insecure logic.
  \item \textbf{Proxy Pattern Challenges:} Contracts using proxy patterns or external libraries pose additional challenges, as dangerous functionality may be located in separate files or storage slots not visible to Slither in isolation~[47].
\end{itemize}
Despite these limitations, our pipeline provides a valuable first-line filter, especially in a space where no review at all is the norm.

\subsection{Broader Implications for NFT Ecosystem Integrity}

The widespread presence of high-risk contract logic calls into question the integrity and sustainability of the current NFT ecosystem. Key concerns include:
\begin{itemize}
  \item \textbf{Reactive Marketplaces:} Marketplaces remain reactive, often delisting collections only after community outcry or exploit~[48].
  \item \textbf{Lack of Developer Incentives:} There is little incentive for developers to relinquish control or follow best practices unless required by platforms.
  \item \textbf{Erosion of Trust:} High-profile scams and rug pulls erode long-term user trust in NFTs as a secure asset class~[44].
\end{itemize}
These trends underscore the need for systemic improvements both technical and policy-based, such as mandatory pre-deployment analysis, risk labeling and stronger community auditing norms.

\subsection{Future Research Directions}

Our study opens multiple avenues for further work:

\begin{enumerate}[label=\arabic*)]
  \item \textbf{Dynamic Analysis Integration:} Future pipelines could combine static analysis with fuzzing or symbolic execution to capture activation conditions and simulate exploit paths~[45].
  \item \textbf{Semantic Labeling:} Machine learning models could be trained on labeled malicious vs.\ benign functions to identify subtle intent-driven code structures~[47].
  \item \textbf{Behavioral Correlation:} Cross-chain event analysis (e.g., abrupt withdrawal patterns, metadata changes) could correlate flagged contracts with actual rug pull events.
  \item \textbf{Cross-Ecosystem Risk Mapping:} Expanding analyses to other chains (Polygon, BNB Chain, Avalanche) may reveal whether these backdoor patterns are chain-specific or part of a broader trend.
\end{enumerate}

The rapid rise of NFTs has brought immense innovation to digital ownership and asset representation but it has also created fertile ground for financial abuse via smart contract backdoors. In this study, we conducted one of the largest static analyses to date on 49,940 verified NFT smart contracts deployed on Ethereum, focusing on hidden logic patterns that facilitate rug pull attacks.

By leveraging Slither and designing a custom heuristic scoring system, we systematically identified and classified high-risk constructs such as selfdestruct, delegatecall, unrestricted mint() functions, and owner-only withdrawals. Our results reveal that more than 22\% of NFT contracts exhibit multiple overlapping vulnerabilities, many of which are unlikely to exist unintentionally. These patterns strongly suggest a recurring design model in which control and withdrawal privileges are deliberately retained by the deployer, in direct contradiction to decentralization principles.

Through our risk classification, co-occurrence analysis, and contract-level case examples, we demonstrate that malicious contract logic is neither rare nor isolated it is embedded across a wide spectrum of deployed NFT projects. Moreover, our findings highlight the inadequacy of current audit practices and the need for stronger pre-deployment screening, automated risk assessment, and marketplace-level safeguards.

While our analysis is static and does not simulate live exploitability, it establishes a scalable and reproducible framework for flagging smart contract backdoors before they are abused in the wild. We hope this work contributes to a more secure, transparent, and accountable NFT ecosystem and that it catalyzes a broader shift toward embedding security as a first-class concern in smart contract development and deployment.

\section{Conclusion}
The rapid rise of NFTs has brought immense innovation to digital ownership and asset representation but it has also created fertile ground for financial abuse via smart contract backdoors. In this study, we conducted one of the largest static analyses to date on 49,940 verified NFT smart contracts deployed on Ethereum, focusing on hidden logic patterns that facilitate rug pull attacks [48].

By leveraging Slither and designing a custom heuristic scoring system, we systematically identified and classified high-risk constructs such as selfdestruct, delegatecall, unrestricted mint() functions, and owner-only withdrawal[49]. Our results reveal that more than 22\% of NFT contracts exhibit multiple overlapping vulnerabilities, many of which are unlikely to exist unintentionally. These patterns strongly suggest a recurring design model in which control and withdrawal privileges are deliberately retained by the deployer, in direct contradiction to decentralization principles [50].

Through our risk classification, co-occurrence analysis, and contract-level case examples, we demonstrate that malicious contract logic is neither rare nor isolated, it is embedded across a wide spectrum of deployed NFT projects [51]. Moreover, our findings highlight the inadequacy of current audit practices and the need for stronger pre-deployment screening, automated risk assessment, and marketplace-level safeguards [52].

While our analysis is static and does not simulate live exploitability, it establishes a scalable and reproducible framework for flagging smart contract backdoors before they are abused in the wild. We hope this work contributes to a more secure, transparent, and accountable NFT ecosystem and that it catalyzes a broader shift toward embedding security as a first-class concern in smart contract development and deployment [53].


\begin{thebibliography}{00}
\bibitem{effective20} Ghaleb, Asem and Pattabiraman, Karthik. "How effective are smart contract analysis tools? evaluating smart contract static analysis tools using bug injection." Proceedings of the 29th ACM SIGSOFT International Symposium on Software Testing and Analysis.
\end{thebibliography}
\end{document}